%% file: paper.tex
\documentclass[10pt]{article}
\usepackage{graphicx}

\usepackage{acm}
\usepackage{fullpage}
\usepackage{url}

\newcommand{\vat}{{\texttt vat}}
\newcommand{\Vat}{{\texttt Vat}}
\usepackage{psfig}
\usepackage{epsfig}
\usepackage{goodfoot}

\date{}

\title{System Support for Bandwidth Management and Content Adaptation
in Internet Applications}

\author{David Andersen, Deepak Bansal, Dorothy Curtis, Srinivasan
Seshan\thanks{Carnegie Mellon University, Pittsburgh, PA;
srini@seshan.org}, Hari Balakrishnan\\{\em M.I.T. Laboratory for
Computer Science}\\{\em Cambridge, MA 02139}\\ \{dga, bansal, dcurtis,
srini, hari\}@lcs.mit.edu}

\begin{document}

\ifpreprint
\makeatletter
\def\headingsize{\@setsize\headingsize{11pt}\viiipt\@viiipt}
\newcommand{\ps@appearsin}{%
  \renewcommand{\@oddhead}{\hfill\begin{minipage}{3.3in}
  \begin{flushright}
\textnormal{\headingsize {\bf }}
  \end{flushright}
  \end{minipage}}%
  \renewcommand{\@evenhead}{}%
  \renewcommand{\@evenfoot}{\hfil\textrm{\thepage}\hfil}%
  \renewcommand{\@oddfoot}{\@evenfoot}}
\makeatother
\thispagestyle{appearsin}
\fi

\maketitle

\begin{abstract}

This paper describes the implementation and evaluation of an operating
system module, the Congestion Manager (CM), which provides integrated
network flow management and exports a convenient programming interface
that allows applications to be notified of, and adapt to, changing
network conditions.  We describe the API by which applications
interface with the CM, and the architectural considerations that
factored into the design.  To evaluate the architecture and API, we
describe our implementations of TCP; a streaming layered audio/video
application; and an interactive audio application using the CM, and
show that they achieve adaptive behavior without incurring much
end-system overhead.  All flows including TCP benefit from the sharing
of congestion information, and applications are able to incorporate
new functionality such as congestion control and adaptive behavior.

\end{abstract}

\begin{sloppypar}

\input{intro}
\input{arch}

\input{apps}

\input{expt}

\input{lexpt}

\input{future}
\input{related}

\input{concl}

\end{sloppypar}

\begin{small}
\bibliography{ref,rfc}
\bibliographystyle{acm}
\end{small}

\end{document}

%% file: intro.tex
\section{Introduction}

The impressive scalability of the Internet infrastructure is in large
part due to a design philosophy that advocates a simple architecture
for the core of the network, with most of the intelligence and state
management implemented in the end systems~\cite{Clark88b}.  The
service model provided by the network substrate is therefore primarily
a ``best-effort'' one, which implies that packets may be lost,
reordered or duplicated, and end-to-end delays may be variable.
Congestion and accompanying packet loss are common in heterogeneous
networks like the Internet because of overload, when demand for router
resources, such as bandwidth and buffer space, exceeds what is
available.  Thus, end systems in the Internet should incorporate
mechanisms for detecting and reacting to network congestion, probing
for spare capacity when the network is uncongested, as well as
managing their available bandwidth effectively.

Previous work has demonstrated that the result of uncontrolled
congestion is a phenomenon commonly called ``congestion
collapse''~\cite{rfc2309,Floyd99a}. Congestion collapse is largely
alleviated today because the popular end-to-end Transmission Control
Protocol (TCP)~\cite{rfc793,Stevens94} incorporates sound congestion
avoidance and control algorithms. However, while TCP does implement
congestion control~\cite{Jacobson88}, many applications including the
Web~\cite{rfc1945,rfc2068} use several logically different streams in
parallel, resulting in multiple concurrent TCP connections between the same
pair of hosts.  As several researchers have
shown~\cite{Balakrishnan98a,Balakrishnan99a,Padmanabhan98,Padmanabhan94,rfc2140}, 
these concurrent connections compete with --
rather than learn from -- each other about network conditions to the
same receiver, and end up being unfair to other applications that use
fewer connections.  The ability to share congestion information
between concurrent flows is therefore a useful feature, one that
promotes cooperation among different flows rather than adverse
competition.

In today's Internet is the increasing number
of applications that do not use TCP as their underlying transport,
because of the constraining reliability and ordering semantics imposed
by its in-order byte-stream abstraction.  Streaming audio and
video~\cite{www-mediaplayer,www-real,Tan99} and customized image
transport protocols %
are significant examples.  Such
applications use custom protocols that run over the User Datagram
Protocol (UDP)~\cite{rfc768}, often without implementing any form of
congestion control.  The unchecked proliferation of such applications
would have a significant adverse effect on the stability of the
network~\cite{Balakrishnan99a,rfc2309,Floyd99a}.

Many Internet applications deliver documents and images or
stream audio and video to end users and are {\em interactive} in
nature.  A simple but useful figure-of-merit for interactive content
delivery is the end-to-end download latency; users typically wait no
more than a few seconds before aborting a transfer if they do not
observe progress. Therefore, it would be beneficial for content
providers to adapt {\em what} they disseminate to the state of the
network, so as not to exceed a threshold latency.  Fortunately, such
content adaptation is possible for most applications. Streaming audio
and video applications typically encode information in a range of
formats corresponding to different encoding (transmission) rates and
degrees of loss resiliency.  Image encoding formats accommodate a
range of qualities to suit a variety of client requirements.

Today, the implementor of an Internet content dissemination
application has a challenging task: for her application to be safe for
widespread Internet deployment, she must either use TCP and suffer the
consequences of its fully-reliable, byte-stream abstraction, or use an
application-specific protocol over UDP.  With the latter option, she
must re-implement congestion control mechanisms, thereby risking errors not just in
the implementation of her protocol, but also in the implementation of
the congestion controller.  Furthermore, neither alternative allows
for sharing congestion information across flows.  Finally, the
common application programming interface (API) classes for network
applications---Berkeley sockets, streams, and
Winsock~\cite{Quinn99}---do not expose any information about the state
of the network to applications in a standard way\footnote{Utilities
like {\tt netstat} and {\tt ifconfig} provide some information about
devices, but not end-to-end performance information that can be used
for adapting content.}. This makes it difficult for applications
running on existing end host operating systems to make an informed
decision, taking network variables into account, during content
adaptation.

\subsection{The Congestion Manager}

Our previous work provided the rationale, initial design, and
simulation of the Congestion Manager, an end-system architecture for
sharing congestion information between multiple concurrent
flows~\cite{Balakrishnan99a}. In this paper, we describe the
implementation and evaluation of the CM in the Linux operating system.
We focus on a version of the CM where the only changes made to the
current IP stack are at the data sender, with feedback about
congestion or successful data receptions being provided by the
receiver CM applications to their sending peers, which communicate
this information to the CM via an API.  We present a summary of the
API used by applications to adapt their transmissions to changing
network conditions, and focus on those elements of the API that
changed in the transition from the simulation to the implementation.

We evaluate the Congestion Manager by posing and answering several key
questions:

{\bf Is its callback interface, used to inform applications of network
state and other events, effective for a diverse set of applications to
adapt without placing a significant burden on developers?}

Because most robust congestion control algorithms rely on receiver
feedback, it is natural to expect that a CM receiver is needed to
inform the CM sender of successful transmissions and packet losses.
However, to facilitate deployment, we have designed our system to take
advantage of the fact that several protocols including TCP and other
applications already incorporate some form of application-specific
feedback, providing the CM with the loss and timing information it
needs to function effectively.

Using the CM API, we implement
several case studies both in and out of the kernel, showing
the applicability of the API to many different application
architectures.
Our implementation of
a layered streaming audio/video application demonstrates that the CM
architecture can be used to implement highly adaptive congestion
controlled applications.  Adaptation via the CM helps these
applications achieve better performance and also be fair to other
flows on the Internet.  

We have also modified a legacy application---the Internet audio tool
{\em vat} from the MASH toolkit~\cite{www-mash}---to use the CM to
perform adaptive real-time delivery.  Since less than one hundred
lines of source code modification was required to CM-enable this
complex application and make it adapt to network conditions, we
believe it demonstrates the ease with which the CM makes applications
adaptive.

{\bf Is the congestion control correct?}

As a trusted kernel module, the CM frees both transport protocols and
applications from the burden of implementing congestion management.
We show that the CM behaves in the same network-friendly manner as TCP
for single flows.  Furthermore, by integrating flow information
between both kernel protocols and user applications, we ensure that an
{\em ensemble} of concurrent flows is not an overly aggressive user of
the network.

{\bf In today's off-the-shelf operating
systems, does the CM place any performance limitations upon
applications?}

We find that our implementation of TCP
(which uses the CM for its congestion control) has essentially the
same performance as standard TCP, with the added benefits of
integrated congestion management across flows, with only 
small (0-3\%) CPU overhead.

In a CM system where no changes are made to the receiver protocol
stack, UDP-based applications must implement a congestion feedback
mechanism, resulting in more overhead compared to the TCP applications.
However, we show that these applications remain viable, and that the
architectural change and API calls reduce worst-case throughput by 0 -
25\%, even for applications that desire fine-grained information about
the network on a per-packet basis.

To our knowledge, this is the first implementation of a general
application-independent system that combines integrated flow
management with a convenient API to enable content adaptation.  The
end-result is that applications achieve the desirable congestion
control properties of long-running TCP connections, together with the
flexibility to adapt data transmissions to prevailing network
conditions.

The rest of this paper is organized as follows.  Section~\ref{s:arch}
describes our system architecture and implementation.
Section~\ref{s:apps} describes how network-adaptive applications can
be engineered using the CM, while Section~\ref{s:expt} presents the
results of several experiments.  In Section~\ref{s:future}, we discuss
some miscellaneous details and open issues in the CM architecture.  We
survey related work in Section~\ref{s:related} and conclude with a
summary in Section~\ref{s:concl}.

%% file: arch.tex
\section{System Architecture and Implementation}
\label{s:arch}

The CM performs two important functions. First, it enables efficient
multiplexing and congestion control by integrating congestion
management across multiple flows. Second, it enables efficient
application adaptation to congestion by exposing its knowledge of
network conditions to applications.  Most of the CM functionality in
our Linux implementation is in-kernel; this choice makes it convenient
to integrate congestion management across both TCP flows and other
user-level protocols, since TCP is implemented in the kernel.

To perform efficient aggregation of congestion information across
concurrent flows, the CM has to identify which flows potentially share
a common bottleneck link {\em en route} to various receivers.  In
general, this is a difficult problem, since it requires an
understanding of the paths taken by different flows.  However, in
today's Internet, all flows destined to the same end host take the
same path in the common case, and we use this group of flows as the
default granularity of flow aggregation\footnote{This is not strictly
true in the presence of network-layer differentiated services.  We
address this issue later in this section and in
Section~\ref{s:future}.}.  We call this group a {\em macroflow}: a
group of flows that share the same congestion state, control
algorithms, and state information in the CM.  Each flow has a sending
application that is responsible for its transmissions; we call this a
{\em CM client}.  CM clients are in-kernel protocols like TCP or
user-space applications.

The CM incorporates a {\em congestion controller} that performs
congestion avoidance and control on a per-macroflow basis.  It uses
a window-based algorithm that mimics TCP's
additive-increase/multiplicative decrease (AIMD) scheme to ensure
fairness to other TCP flows on the Internet.  However, the modularity
provided by the CM encourages experimentation with other non-AIMD
schemes that may be better suited to specific data types such as audio
or video.

While the congestion controller determines what the current window
(rate) ought to be for each macroflow, a {\em scheduler} decides how
this is apportioned among the constituent flows.  Currently, our
implementation uses a standard unweighted round-robin scheduler.

\begin{figure}[t]
\begin{center}
\leavevmode
\hbox{%
\psfig{figure=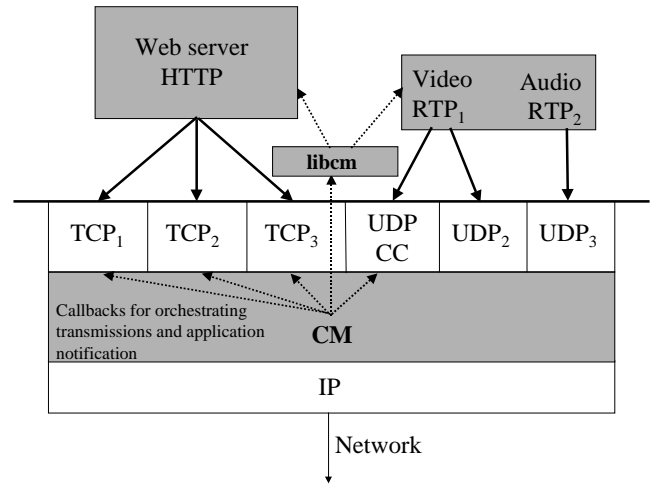,width=0.8\columnwidth,angle=270}}
\caption{Architecture of the congestion manager at the data sender,
showing the CM library and the CM.  The dotted arrows show callbacks,
and solid lines show the datapath.
UDP-CC is a congestion-controlled UDP socket implemented using the CM.}
\label{f:cmarch}
\end{center}
\end{figure}

In-kernel CM clients such as a TCP sender use CM function calls to
transmit data and learn about network conditions and events.  In
contrast, user-space clients interact with the CM using a portable,
platform-independent API described in Section~\ref{s:api}.  A
platform-dependent CM library, {\tt libcm}, is responsible for
interfacing between the kernel and these clients, and is described in
Section~\ref{s:libcm}.  These components are shown in
Figure~\ref{f:cmarch}.

When a client opens a CM-enabled socket, the CM allocates a flow to it
and assigns the flow to the appropriate macroflow based on its
destination.  The client initiates data transmission by requesting
permission to send data.  At some point in the future depending on the
available rate, the CM issues a callback permitting the client to send
data.  The client then transmits data, and tells the CM it has done
so.  When the client receives feedback from the receiver about its
past transmissions, it notifies the CM about these and continues.

When a client makes a request to send on a flow, the scheduler checks
whether the corresponding macroflow's window is open.  If so, the
request is granted and the client notified, upon which it may send
some data.  Whenever any data is transmitted, the sender's IP layer
notifies the CM, allowing it to ``charge'' the transmission to the
appropriate macroflow.  When the client receives feedback from its
remote counterpart, it informs the CM of the loss rate, number of
bytes transmitted correctly, and the observed round trip time.  On a
successful transmission, the CM opens up the window according to its
congestion management algorithm and grants the next, if any, pending
request on a flow associated with this macroflow.  The scheduler also
has a timer-driven component to perform background tasks and error
handling.

\subsection{CM API}
\label{s:api}

The CM API is specified as a set of functions and callbacks that a
client uses to interface with the CM.  It specifies functions for
managing state, for performing data transmissions, for 
applications to inform the CM of losses, for querying the CM about
network state, and for constructing and splitting macroflows if the
default per-destination aggregation is unsuitable for an application.
The CM API is discussed in detail in ~\cite{Balakrishnan99a},
which presents the design rationale for the Congestion Manager.
Here we provide an overview of the API and a discussion of those
features which changed during the transition from simulation
to implementation.

\subsubsection{State management}
All CM applications call {\tt cm\_open()} before using the CM, passing
the source and destination addresses and transport-layer port numbers,
in the form of a {\tt struct sockaddr}.
The original CM API required only a destination address, but the
source address specification was necesary to handle multihomed
hosts.  {\tt cm\_open} returns a flow identifier
({\tt cm\_flowid}), which is used as a handle for all future CM calls.
Applications may call {\tt cm\_mtu(cm\_flowid)} to obtain the
maximum transmission unit to a destination.
When a flow terminates, the application should call 
{\tt cm\_close(cm\_flowid)}.

\subsubsection{Data transmission}
There are three ways in which an application can use the CM to
transmit data.  These allow a variety of adaptation strategies,
depending on the nature of the client application and its software
structure.

\begin{enumerate}
\item [(i)] {\bf Buffered send.}  This API uses a conventional
{\tt write()} or {\tt sendto()} call, but the resulting data transmission is
paced by the Congestion Manager.  We use this to implement a generic
congestion-controlled UDP socket (without content
adaptation), useful for bulk transmissions that
do not require TCP-style reliability or fine-grained control over what
data gets sent at a given point in time.

\item [(ii)] {\bf Request/callback.}  This is the preferred mode of
communication for adaptive senders that are based on the ALF
(Application-Level Framing~\cite{Clark90})
principle.  Here, the client does not send data via the CM;
rather, it calls {\tt cm\_request(cm\_flowid)} and expects a
notification via the {\tt cmapp\_send(cm\_flowid)} callback
when this request is granted by the CM,
at which time the client transmits its data.
This approach puts the sender in firm control of deciding
what to transmit at a given time, and allows the sender to
adapt to sudden changes in network performance, which is hard to do in
a conventional buffered transmission API.  The client callback is a
grant for the flow to send up to MTU bytes of data.  Each call
to {\tt cm\_request()} is an implicit request
for sending up to MTU bytes, which simplifies the
internal implementation of the CM.
This API is ideally
suited for an implementation of TCP, since it needs to make a decision
at each stage about whether to retransmit a segment or send a new one.
In the implementation, the {\tt cmapp\_send} callback now
provides the client with the ID of the flow that may transmit.
To allow for client programming flexibility, the client
may now specify its callback function via {\tt cm\_register\_send()}.

\item [(iii)] {\bf Rate callback.}  A self-timed application 
transmitting on a
fixed schedule may receive callbacks from the CM notifying it when the
parameters of its communication channel have changed, so that it can
change the frequency of its timer loop or its packet size.
The CM informs the client of the rate, round-trip time, and packet
loss rate for a flow via the {\tt cmapp\_update()} callback.
During implementation, we added a registration function,
{\tt cm\_register\_update()} to select the rate callback
function, and the {\tt cm\_thresh(down,up)} function:
If the rate reduces by a factor of {\tt down} or increases by a
factor of {\tt up}, the CM calls {\tt cmapp\_update()}.  This
transmission API is ideally suited for streaming layered audio and
video applications.
\end{enumerate}

\subsubsection{Application notifications}
One of the goals of our work was to investigate a CM implementation
that requires no changes at the receiver.  Performing
congestion management requires feedback about transmissions:
TCP provides this feedback automatically; some UDP
applications may need to be modified to do so, but without
any system-wide changes.
Senders must then inform the CM about the
number of sent and received packets, type of congestion loss if any,
and a round-trip time sample
using the 
{\tt cm\_update(
cm\_flowid, nsent, nrecd, lossmode, rtt)} function.
The CM distinguishes between
``persistent'' congestion as would occur on a TCP timeout, versus
``transient'' congestion when only one packet in a window is lost.  It
also allows congestion to be notified using Explicit Congestion
Notification (ECN)~\cite{rfc2481}, which uses packet markings rather than
drops to infer congestion.

To perform accurate bookkeeping of the congestion window and
outstanding bytes for a macroflow, the CM needs to know of each
successful transmission from the host.  Rather than encumber clients
with reporting this information, we modify the IP output routine to call
{\tt cm\_notify(cm\_flowid, nsent)} on each transmission.  (The IP
layer obtains the {\tt cm\_flowid} using a well-defined CM interface
that takes the flow parameters (addresses, ports, protocol field) as
arguments.)  However, if a client decides not to transmit any data
upon a {\tt cmapp\_send()} callback invocation, it is expected to call
{\tt cm\_notify(dst, 0)} to allow the CM to permit some other flows on
the macroflow to transmit data.

\subsubsection{Querying}

If a client wishes to learn about its (per-flow) available bandwidth
and round-trip time, it can use the {\tt cm\_query()} call, which
returns these quantities.  This is especially useful at the beginning
of a stream when clients can make an informed decision about the data
encoding to transmit (e.g., a large color or smaller grey-scale
image).

\subsection{{\tt libcm}: The CM library}
\label{s:libcm}

The CM library provides users with the convenience of a
callback-based API while separating them from the details of how the
kernel to user callbacks are implemented.  While direct function
callbacks are convenient and efficient in the same address space, as
is the case when the kernel TCP is a client of the CM,
callbacks from the kernel to user code in
conventional operating systems are more difficult.  A key decision in
the implementation of {\tt libcm} was choosing a kernel/user
interface that maximizes portability, and minimizes both performance
overhead and the difficulty of integration with existing applications.
The resulting \emph{internal} interface between {\tt libcm} and
the kernel is:

\begin{enumerate}
 \item {\tt select()} on a single per-application CM control socket.
       The write bit indicates that a flow may send data, and the
       exception bit indicates that network conditions have changed.

\item Perform an {\tt ioctl} to extract a list of all flow IDs
       which may send, or to receive the current network conditions
       for a flow.
\end{enumerate}

Note that client programs of the CM do not see this interface;  they
see only the standard {\tt cm\_*} functions provided by
{\tt libcm}.  The use of sockets or signals does change the way
the application's event handling loop interacts with {\tt libcm};
after passing the socket into {\tt libcm}, the library performs
the appropriate {\tt ioctl}s and then calls back into the
application.

\subsubsection{Implementation alternatives}

We considered a number of mechanisms with which to implement {\tt
libcm}.  In this section, we discuss our reasons for choosing the
{\texttt control-socket+select+ioctl} approach.

While much research has focused on reducing the cost of crossing the
user/kernel boundary (extensible kernels in SPIN
\cite{Bershad+:sosp15}, fast, generic IPC in
Mach~\cite{Barrera91}, etc.) many conventional operating systems remain
limited to more primitive methods for kernel-to-user notification,
each with their own advantages and disadvantages.  While functionality
like the Mach port set-based IPC would be ideal for
our purposes, pragmatically
we considered four common mechanisms for kernel
to user communication:  Signals, system calls, semaphores, and
sockets.  A discussion of the merits of each follows.

{\bf Signals} have several immediate drawbacks.  First, if the CM were to
appropriate an existing signal for its own use, it might conflict with
an application using the same signal.  Avoiding this conflict would
require the standardization of a new signal type, a process both slow
and of questionable value, given the existence of better alternatives.
Second, the cost to an application to receive a signal is relatively
high, and some legacy applications may not be signal-safe.  While the
new POSIX 1003.1b~\cite{POSIX-1b} soft realtime signals allow
delivering a 32-bit quantity with a signal, applications would need to
follow up a signal with a system call to obtain all of the information
the kernel wished to deliver, since multiple flows may become ready at
once.  For these reasons, we consider mandating the use of signals the
wrong course for implementing the kernel to user callbacks.  However, we
provide an \emph{option} for processes to receive a {\tt SIGIO}
when their control socket status changes, akin to POSIX
asynchronous I/O.

{\bf System calls} that block do not integrate well with applications that
already have their own event loop, since without polling, applications
cannot wait on the results of multiple system calls.  A system call is
able to return immediately with the data the user needs, but the
impediments it poses to application integration are large.  System
calls would work well in a threaded environment, but this presupposes
threading support, and the select-based mechanism we describe below
can be used in a threaded system without major additional overhead.

{\bf Semaphores} suffer from the immediate drawback that they are not
commonly used in network applications.  For an application that uses
{\tt semop} on an array of semaphores as its event loop, a CM
semaphore might be the best implementation avenue, for many of the
same reasons that we chose sockets for network-adaptive applications.
However, most network applications use socket sets instead of
semaphore sets, and sockets have a few other benefits, which we
discuss next.

{\bf Sockets} provide a well-defined and flexible interface for applications
in the form of the {\tt select()} system call, though they have a
downside similar to that of signals: an application wishing to receive
a notification via a socket in a non-blocking manner must {\tt
select()} on the socket, and then perform a system call to obtain data
from the socket.  However, a select-based interface meshes well with
many network applications that already have a select-loop based
architecture.  Utilizing a control socket also helps restrict the code
changes caused by the CM to the networking stack.

Finally, we decided to use a single control socket instead of one
control socket per flow to avoid unnecessary overhead in applications
with large numbers of open socket descriptors, such as {\tt
select()}-based webservers and caches.  Because some aspects of select
scale linearly with the number of descriptors, and many operating
systems have limits on the number of open descriptors, we deemed
doubling the socket load for high-performance network applications a
bad idea.

\subsubsection{Extracting data from the socket}

Select provides notification
that ``some event'' has occured.
In theory, 7 different events could
be sent by abusing the read, write, and exception bits,
but applications need to extract more
information than this.
The CM provides two types of callbacks.  Generally speaking,
the first is a ``permission to send'' callback for a
particular flow.  To maintain
even distribution of bandwidth between flows, a loose ordering should be preserved with these
messages, but exact ordering is unimportant provided
no flows are ignored until the application receives
further updates (thereby starving the flows).  If
multiple permission notifications occur, the application
should receive \emph{all} of them so it can send data
on all available flows.
The second callback 
is a ``status changed'' notification.  If multiple status
changes occur before the application obtains this data
from the kernel, then only the \emph{current} status
matters.

The weak ordering and lack of history
prompted us to choose an {\tt ioctl}-based
query instead of a {\tt read} or message queue
interface, minimizing the state that must be maintained
in the kernel.  Status updates simply return the
current CM-maintained network state estimate, and
``who can send'' queries perform a select-like
operation on the flows maintained by the kernel,
requiring no extra state,
instead of a potentially expensive per-process message queue
or data stream.  Returning all available flows
has an added benefit of 
reducing the number of system calls that must
be made if several flows become ready
simultaneously.

%% file: apps.tex
\section{Engineering Network-adaptive Applications}
\label{s:apps}

In this section, we describe several different
classes of applications, and describe the ways
those applications can make use of the CM.  We
explore two in-kernel clients, and several
user-space data server programs, and examine
the task of integrating each with the CM.

\input{sftwr}

The remainder of this section describes how particular clients
use different CM APIs,
from the low-bandwidth {\tt vat} audio application, to the
performance-critical kernel TCP implementation.
Note that all UDP-based clients must implement
application level data acknowledgements in
order to make use of the CM.

\subsection{TCP}
We implemented TCP as an in-kernel CM client.
TCP/CM offloads all congestion control to the CM,
while retaining all other TCP functionality
(connection establishment and termination, loss recovery
and protocol state handling).
TCP uses the request/callback API as
low-overhead direct function calls in the same protection domain.
This gives TCP the tight control it needs over
packet scheduling.
For example, while the arrival of a new acknowledgement
typically causes TCP to transmit new data, the arrival
of three duplicate ACKs causes TCP to retransmit
an old packet.

{\bf Connection creation}.
When TCP creates a new connection 
via either {\tt accept} (inbound) or {\tt connect} (outbound),
it calls
{\tt cm\_open()} to associate the TCP connection
with a CM flow.
Thereafter, the pacing of outgoing data 
on this connection is controlled by the CM.
When application data becomes available, after performing 
all the non-congestion-related checks (e.g., the Nagle
algorithm~\cite{Stevens94}, etc.)
data is queued and 
{\tt cm\_request()} is called for the flow. When the CM scheduler
schedules the flow for transmission, the  
{\tt cmapp\_send()} routine  for TCP is called. The {\tt cmapp\_send()}
for TCP  
transmits any
retransmission from the retransmission queue. Otherwise, it
transmits the data present in the transmit socket
buffer by sending up to one maximum segment size of data per
call. Finally, the IP output routine calls {\tt
cm\_notify()} when the data is actually sent out.

{\bf TCP input}.
The TCP input routines now feedback to the CM.
Round trip time (RTT) sample collection is done as usual using either RFC
1323 timestamps~\cite{rfc1323} or Karn's algorithm~\cite{Karn91}
and is passed to CM via {\tt
cm\_update()}.  The smoothed estimates of the RTT  ({\tt
srtt}) and round-trip time deviation are calculated by the CM, which
can now obtain a better average by combining samples from
different connections to the same receiver.  This is available to
each TCP connection via {\tt cm\_query()}, and is useful in loss recovery.

{\bf Data acknowledgements}.
On arrival of an ACK for new data, 
the TCP sender  calls
{\tt cm\_update()}
to inform the CM of a successful transmission.
Duplicate acknowledgements cause TCP to
check its dupack count ({\tt dup\_acks}).
If {\tt dup\_acks} $< 3$, then TCP does nothing. 
If {\tt dup\_acks $== 3$}, then TCP assumes a simple,
congestion-caused packet loss,
and calls {\tt cm\_update} 
to inform the
CM.
TCP also enqueues a retransmission of the lost segment and calls
{\tt cm\_request()}. If {\tt dup\_acks} $> 3$, TCP assumes that a
segment reached the receiver and caused this ACK to be sent. It
therefore calls {\tt cm\_update()}.
Unlike duplicate ACKs, the expiration of
the TCP retransmission timer notifies the sender of
a more serious batch of losses, so it calls {\tt cm\_update}
with the CM\_LOST\_FEEDBACK option set to signify the occurrence of
persistent congestion to the CM.  TCP also enqueues a retransmission
of the lost segment and calls {\tt cm\_request()}.

{\bf TCP/CM Implementation}.
The integration of TCP and the CM required less than 100 lines
of changes to the existing TCP code, demonstrating
both the flexibility of the CM API and the
low programmer overhead of implementing
a complex protocol with the Congestion Manager.

\subsection{Congestion-controlled UDP sockets}

The CM also provides congestion-controlled
UDP sockets.
They provide the same functionality as standard
Berkeley UDP sockets, but instead of
immediately sending the data from the kernel packet queue 
to lower layers for transmission, 
the buffered socket implementation
schedules its packet output via CM callbacks.
When a CM UDP socket is created, it is bound
to a particular flow.
When data enters the packet queue, the kernel calls
{\tt cm\_request()} on the flow associated with the socket.
When the CM schedules this flow for transmission,
it calls
{\tt udp\_ccappsend()} 
in the CM UDP module.  This function transmits
one MTU from the packet queue,
and requests another callback if packets remain.
The in-kernel implementation of the CM UDP API
adds no data copies or queue structures,
and supports all standard UDP options.
Modifying existing applications to
use this API requires only providing feedback to the
CM, and setting a socket option on the socket.

A typical client of the CM UDP sockets will behave as follows,
after its usual network socket initialization:

\begin{verbatim}
  flow = cm_open(dst, port)
  setsockopt(flow, ..., CM_BUF)
  loop:
    <send data on flow>
    <receive data acknowledgements>
    cm_update(flow, sent, received, ...)
\end{verbatim}

\subsection{Streaming Layered Audio and Video}

Streaming layered audio or video applications that
have a number of discrete rates at which they can transmit data
are well-served by the CM rate callbacks.  Instead of
requiring a comparatively expensive notification for
each transmission, these applications are instead notified
only in the rare event that their network conditions
change significantly.  Layered applications open their usual UDP socket,
and call {\tt cm\_open()} to obtain a control socket.
They operate in their own clocked event
loop while listening for status changes on 
either their control socket or via a SIGIO signal.
They use {\tt cm\_thresh()} to inform the CM about network changes for which
they should receive callbacks.

\subsection{Real-time Adaptive Applications}

Applications that desire last-minute control over their
data transmission
(i.e. those that do not want any buffering inside the kernel) use
the request callback API provided
by the CM. When given permission to transmit via the 
{\tt cmapp\_send()} callback from the CM,
they may use {\tt cm\_query()} to discover the
current network conditions and adapt
their content based on that. Other servers may
simply wish to send the most up-to-date content possible,
and so will defer their data collection until they know they
can send it.  The rough sequence of CM calls that are made
to achieve this in the application are:

\begin{verbatim}
 flow = cm_open(dst)
 cm_request(flow)
 <receive cmapp_send() callback from libcm>
 cm_query(flow, ...)
 <send data>
 <receive data acks>
 cm_update(flow, sent, lost, ...)
\end{verbatim}

Other options exist for applications that
wish to exploit the unique nature of their
network utilization to reduce the overhead
of using the services of the Congestion Manager.
We discuss one such option below in the manner
in which we adapted the {\tt vat} interactive
audio application to use the CM.

\subsection{Interactive Real-time Audio}

The \vat\ application provides a constant bit-rate source
of interactive audio.  Its inability to downsample its
audio reduces the avenues it has available for bandwidth
adaptation.  Therefore, the best way to make \vat\
behave in a network-friendly and backwards compatible
manner is to preemptively drop packets to match the
available network bandwidth.
There are, of course, complications.
Network applications experience
two types of variation in available network bandwidth: long
term variations due to changes in actual bandwidth, and short
term variations due to the probing mechanisms of the congestion
control algorithm.
Short-term variation
is typically dealt with by buffering.
Unfortunately, buffering, especially FIFO buffering
with drop-tail behavior, the de-facto standard for
kernel buffers and network router buffers,
can result in long delay and significant delay variation,
both of which are detrimental to vat's audio quality.
\Vat\ , therefore, needs to act like an ALF application,
managing its own buffer space with drop-from-head
behavior when the queue is full.

\begin{figure}
  \begin{center}
    \includegraphics[width=0.65\columnwidth]{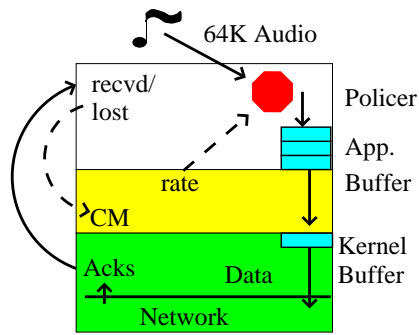}
  \end{center}
  \caption{The adaptive vat architecture}
  \label{figure:arch}
\end{figure}

The resulting architecture is detailed
in figure \ref{figure:arch}.
The input audio stream is first
sent to a policer, which provides
long-term adaptation via preemptive packet dropping.
The policer outputs into the application
level buffer, which can be configured
in various sizes and drop policies.
This buffer feeds into the kernel buffer
on-demand as packets are available
for transmission.

%% file: sftwr.tex
\subsection{Software Architecture Issues}

Typical network applications fall into one
of several categories:

\begin{itemize}
\item Data-driven:  Applications that transmit prespecified
data, such as a single file, then exit.
\item Synchronous event-driven:  Self-timed data delivery
servers, like streaming audio servers.
\item Asynchronous event-driven:  File servers (http, ftp)
and other network-clocked applications.
\end{itemize}

The CM library provides several options
for adaptive applications that wish to make use of
its services:

\begin{enumerate}
\item Data-driven applications may use the buffered %
API to efficiently pace their data transmissions.
\item An application may
operate in an entirely callback-based manner by allowing
{\tt libcm} to provide its own event loop, calling into
the application when flows are ready.  This is most
useful for applications coded with the CM in mind.

\item Signal-driven applications may request a
SIGIO notification from the CM when an event occurs.

\item Applications with select-based event
loops can simply add the CM control socket into
their select set, and call the {\tt libcm} dispatcher
when the socket is ready.  Rate-clocked applications
(or polling-based applications)
can perform a similar nonblocking select test on
the descriptor when they awaken to send data,
or, if they {\tt sleep}, can replace the sleep
with a timed blocking {\tt select} call.

\item Applications may poll the CM on their own
schedule.

\end{enumerate}

%% file: expt.tex
\section{Evaluation}
\label{s:expt}
This section describes several experiments that quantify the costs and
benefits of our CM implementation.  Our experiments show that
using the Congestion Manager in the kernel has minimal costs,
and that even the worst-case overhead 
of the request/callback user-space API is
acceptably small.

The tests were performed on the Utah Network Testbed
\cite{lepreau:sigcomm99-wip}
using 600MHz Intel Pentium III processors, 128MB PC100 ECC SDRAM,
and Intel EtherExpress Pro/100B Ethernet cards, connected via
100Mbps Ethernet through an Intel Express 510T
switch, with Dummynet channel simulation.  CM tests were run
on Linux 2.2.9, with Linux and FreeBSD clients.

To ensure the proper behavior of a flow, the congestion control
algorithm must behave in a ``TCP-compatible''~\cite{rfc2309} manner.
The CM implements a TCP-style window-based AIMD algorithm with slow
start.  It shares bandwidth between eligible flows in a round-robin
manner with equal weights on the flows.

\begin{figure}[t]
\begin{center}
\leavevmode
\hbox{%
\psfig{figure=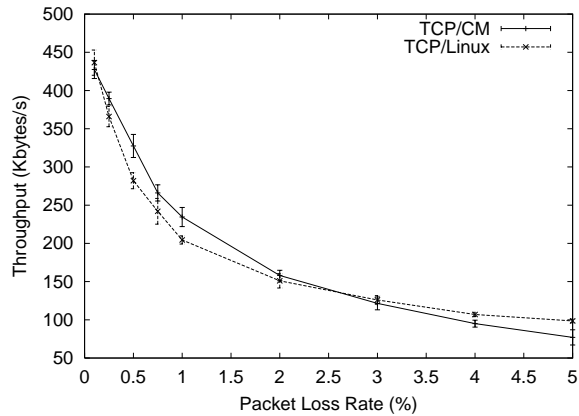,width=0.95\columnwidth}}
\caption{Comparing throughput vs. loss for TCP/CM and TCP/Linux.
Rates are for
a 10Mbps link with a 60ms RTT.}
\label{f:tcp-compare}
\end{center}
\end{figure}

Figure \ref{f:tcp-compare} shows the throughput achieved
by the Linux TCP implementation (TCP/Linux) and TCP with
congestion control performed by the CM
(TCP/CM).
The linux kernel against which we compare 
has two algorithmic differences from the Congestion Manager:
It starts its initial window at 2 packets, and it assumes that
each ACK is for a full MTU.  The Congestion Manager
instead performs byte-counting for its AIMD algorithm.
The first issue is Linux-specific, and
the last is a feature of the CM.

\subsection{Kernel Overhead}
To measure the kernel overhead, we measured the CPU and throughput
differences between the optimized TCP/Linux and TCP/CM.
The midrange machines used
in our test environment are sufficiently powerful to saturate
a 100Mbps Ethernet with TCP traffic.

There are two components to the overhead imposed by the congestion
manager: The cost of performing accounting as
data is exchanged on a connection, and a
one-time connection setup cost for creating CM data structures.
A microbenchmark of
the connection establishment time of a TCP/CM vs.
TCP/Linux indicates that there is no appreciable
difference in connection setup times.

\begin{figure}[t]
\begin{center}
\leavevmode
\hbox{%
\psfig{figure=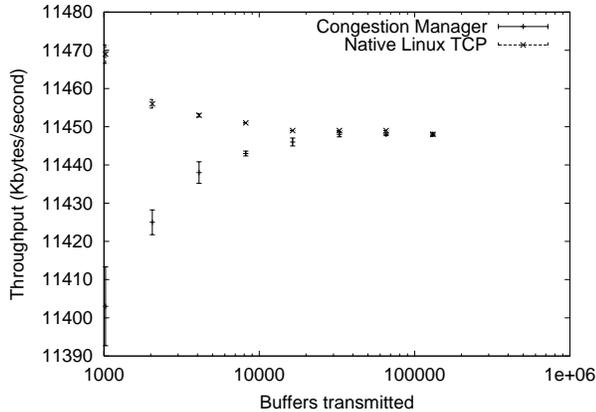,width=0.95\columnwidth}}
\caption{100Mbps TCP throughput comparison.  Note
that the absolute difference in the worst case
between the Congestion Manager and the native TCP
is only 0.5\% and that the Y axis begins
at 11 megabytes per second.}
\label{f:tcp-throughput}
\end{center}
\end{figure}

We used long (megabytes to gigabytes) connections with the {\tt ttcp}
utility to determine the long-term costs imposed by the congestion
manager.  The impact of the CM on extremely long term throughput was
negligible: in a 1 gigabyte transfer, the congestion manager achieved
identical performance (91.6 Mbps) as native Linux.  On shorter runs,
the throughput of the CM diverged slightly from that of Linux, but
only by 0.5\%.  The throughput rates are shown in figure
\ref{f:tcp-throughput}.  The difference is due to the CM using
an initial window of 1 MTU and Linux using 2 MTU, not CPU overhead.

\begin{figure}[t]
\begin{center}
\leavevmode
\hbox{%
\psfig{figure=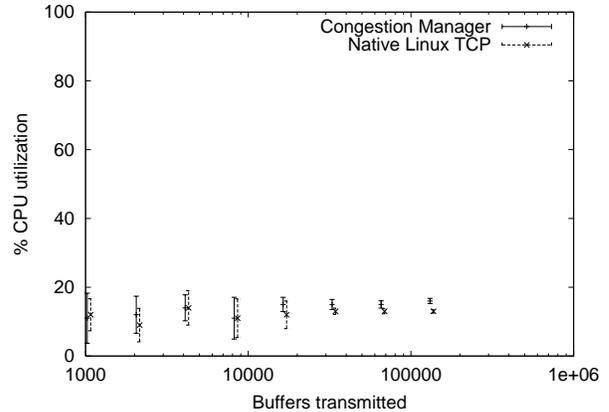,width=0.95\columnwidth}}
\caption{CPU overhead comparison between TCP/Linux and
TCP/CM.  For long connections, the CPU overhead converges to slightly
under 1\% for the unoptimized implementation of the CM.}
\label{figure:tcp-overhead}
\end{center}
\end{figure}

Because both implementations are able to saturate the network
connection, we looked at the CPU utilization during
these transmissions to determine the steady-state overhead
imposed by the Congestion Manager.  In figure
\ref{figure:tcp-overhead}
we see that the CPU difference between TCP/Linux and TCP/CM converges
to slightly less than 1\%.

\subsection{User-space API Overhead}

The overhead incurred by our adaptation API occurs
primarily because the applications must process their
ACKs in user-space instead of in the kernel.  Therefore,
these programs incur extra data copies and user/kernel
boundary crossings.
To quantify this overhead, our
test programs sent packets of specified sizes on a UDP
socket, and waited for acknowledgement packets from the
server.  We compare these programs to a webserver-like
TCP client which sendt data to the server, and performed
a {\tt select()} on its socket to determine if the server
has sent any data back.  To facilitate comparison,
we disabled delayed ACKs for the one TCP test to ensure that
our packet counts were identical.

\begin{figure}[t]
\begin{center}
\leavevmode
\hbox{%
\psfig{figure=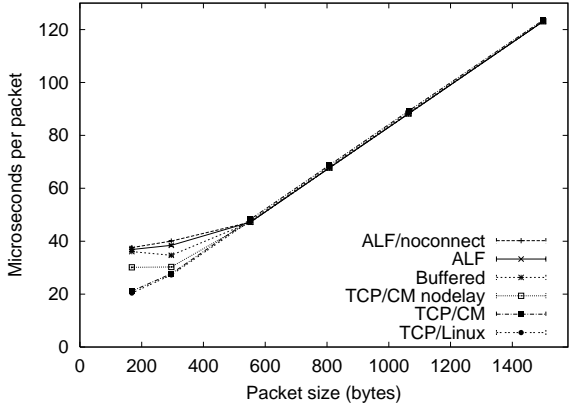,width=0.95\columnwidth}}
\caption{API throughput comparison on a 100Mbps link.
The worst-case throughput reduction incurred by the CM is
25\% from TCP/CM nodelay to ALF/noconnect.}

\label{f:api-compare}
\end{center}
\end{figure}

Figure \ref{f:api-compare} shows the wall-clock time required
to send and process the acknowledgement for a packet, based on
transmitting 200,000 packets.  For comparison, we include
TCP statistics as well, where the TCP programs set the
maximum segment size to achieve identical network
performance.  
The ``nodelay'' variant is TCP without delayed acks.
The tests were run on a 100Mbps network on which no losses occured.

\begin{table}
\begin{tabular}{|l|l|}
\hline
ALF/noconnect  &  1 {\tt cm\_notify}  (ioctl)\\
\hline
ALF            &  1 {\tt cm\_request} (ioctl)\\
               &  1 extra socket\\
\hline
Buffered       &  1 {\tt recv}, 2 {\tt gettimeofday}\\
\hline
TCP/CM         & --baseline--\\
\hline
\end{tabular}
\caption{Cumulative sources of overhead for different
APIs using the Congestion Manager relative to
sending data with TCP.}
\label{table:overheads}
\end{table}

Table \ref{table:overheads} breaks down the sources
of overhead for using the different APIs.  Using the
CM with UDP requires that applications compute the
round-trip-time (RTT) of their packets, requiring a
system call to {\tt gettimeofday}, and requires that
they process their ACKs in user-space, requiring a
system call to {\tt recv} and the accompanying data
copy into their address space.  The ALF API further
requires that the application obtain an additional control
socket and select upon it, and that it make an explicit
call to {\tt cm\_request} before transmitting data.
Finally, if the kernel is unable to determine the
flow to which to charge the transmission, as with
an unconnected UDP socket, the application must
explicitly call {\tt cm\_notify}

These test cases represent the worst-case behavior
of serving a single high-bandwidth client, because
no aggregation of requests to the CM may occur.
The CM programs can achieve similar reductions
in processing time by using delayed acks, so 
the real API overhead can be determined by comparing
the ALF/noconnect case to the TCP/CM case.  For 168 byte
packets, ALF/noconnect results in a 25\% reduction
in throughput relative to TCP without delayed ACKs.

%% file: lexpt.tex
\subsection{Benefits of Sharing}
\label{section:sharing-benefits}

\begin{figure}[t]
\begin{center}
\leavevmode
\hbox{%
\psfig{figure=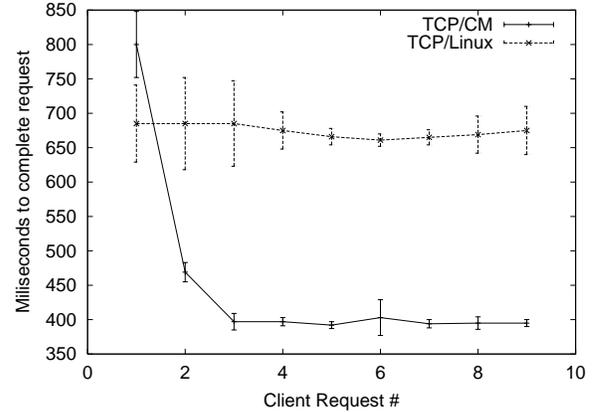,width=0.95\columnwidth}}
\caption{Sharing TCP state:  The client requests the same file 9 times with a
500ms delay between request initiations.
By sharing congestion information and avoiding slow-start, the
CM-enabled server is able to provide faster service for subsequent
requests, despite a smaller initial congestion window.}
\label{f:sharing}
\end{center}
\end{figure}

One benefit of integrating congestion information with the CM is
immediately clear.  A client that sequentially fetches files from a
webserver with a new TCP connection each time loses its prior
congestion information, but with concurrent connections with the CM,
the server is able to use this information to start subsequent
connections with more accurate congestion windows.  Figure
\ref{f:sharing} shows a test we performed across the vBNS between MIT
and the University of Utah, where an unmodified (non-CM) client
performed 9 retrievals of the same 128k file with a 500ms delay
between retrievals, resulting in a 40\% improvement in the transfer
time for the later requests.  (Other file sizes and delays yield
similar results, so long as they overlap.  The benefits are
comparatively greater for smaller files).
The CM requires an additional RTT (~75ms) for the first transfer, because
Linux sets its initial congestion window to 2 MTUs instead of 1.
This pattern of multiple connections is still quite
common in webservers despite the adoption of persistent connections:
Many browsers open 4 concurrent connections to a server,
and many client/server combinations 
do not support persistent connections.
Persistent connections
\cite{Padmanabhan94} provide similar performance
benefits, but suffer from their own drawbacks, which we discuss
in section~\ref{s:related}.

\subsection{Adaptive Applications}
\label{ss:lexpt} 

In this section, we demonstrate some of the network adaptive behaviors
enabled by the CM.

\begin{figure}[t]
\begin{center}
\leavevmode
\hbox{%
\psfig{figure=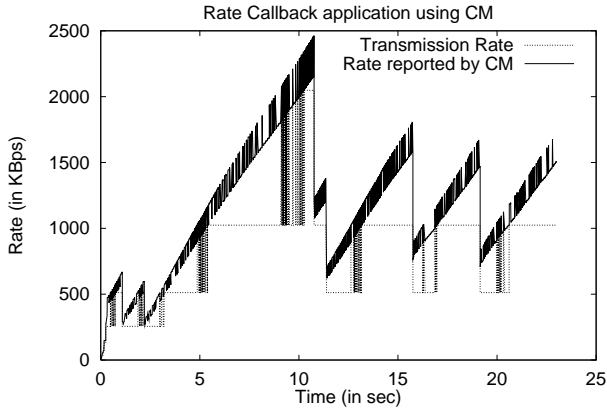,width=0.95\columnwidth}}
\caption{Bandwidth perceived by an adaptive layered application using
the request callback (ALF) API.}
\label{f:lapp-request}
\end{center}
\end{figure}

As noted earlier, applications that require tight control over data
scheduling use the request/callback (ALF) API, and are notified by the
CM as soon as they can transmit data.  The behavior of an adaptive
layering application run across the vBNS
using this API is shown in
figure~\ref{f:lapp-request}.  This application chooses a layer to
transmit based upon the current rate, but sends packets as rapidly as
possible to allow its client to buffer more data.  We see that the CM
is able to provide sufficient information to the application to allow
it to adapt properly to the network conditions.

\begin{figure}[t]
\begin{center}
\leavevmode
\hbox{%
\psfig{figure=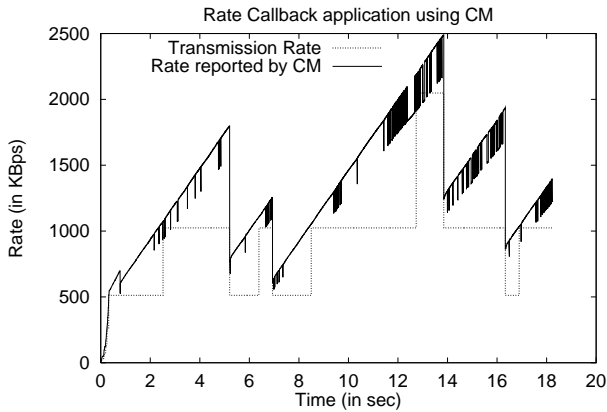,width=0.95\columnwidth}}
\caption{Bandwidth perceived by an adaptive layered application using
the rate callback API.}
\label{f:lapp-rate}
\end{center}
\end{figure}

For self-clocked applications that base their transmitted
data upon the bandwidth to the client (such as conventional
layered audio servers), the CM rate callback mechanism
provides a low-overhead mechanism for adaptation, and allows
clients to specify threshholds for the notification callbacks.
Figure~\ref{f:lapp-rate}
shows application adaptation using rate callbacks for a connection
between MIT and the University of Utah.
Here, the application decides which of the four layers it should send
based on notifications from the CM about rate changes.

From figures~\ref{f:lapp-request} and \ref{f:lapp-rate},
we see from the increased oscillation rate in the transmitted
layer that the ALF application is more responsive
to smaller changes in available bandwidth, whereas the
rate callback application relies occasionally on short-term kernel
buffering for smoothing.  There is an
overhead vs. functionality trade-off in the
decision of which API to use, given the higher overhead of
the ALF API, but applications face a more important
decision about the behavior they desire.

\begin{figure}[t]
\begin{center}
\leavevmode
\hbox{%
\psfig{figure=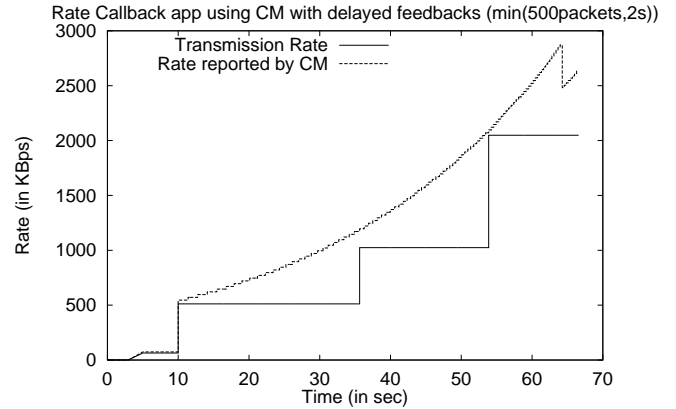,width=0.95\columnwidth}}
\caption{Adaptive layered application using rate callback API with
delayed feedback}
\label{f:lfeedback500}
\end{center}
\end{figure}

Some applications may be concerned about the overhead from
receiver feedback.
To mitigate this, an application may delay sending feedback;  we
see this in a minor and inflexible way with TCP delayed acks.
In figure~\ref{f:lfeedback500}, we see that delaying feedback 
to the CM causes burstiness in the reported bandwidth.
Here, the feedback by the
receiver was delayed by $min($500 acks, 2000ms$)$.
The initial slow start is delayed by 2s waiting for the application,
then the update causes a large rate change.
Once the pipe is sufficiently full,
500 acks come relatively rapidly, and the normal, though bursty, non-timeout
behavior resumes.

%% file: future.tex
\section{Discussion}
\label{s:future}

We have shown several benefits of integrated flow management and the
adaptation API, and have explored the design features that make the
API easy to use.  This section describes an optimization useful for
busy servers, and discusses some drawbacks and limitations of the current CM
architecture. 

{\bf Optimizations.}  Servers with large numbers of concurrent clients
are often very sensitive to the overhead caused by multiple kernel
boundary crossings.  To reduce this overhead, we can batch several
sockets into the same {\tt cm\_request} call with the {\tt
cm\_bulk\_request} call, and likewise for {\tt query}, {\tt
notify}, and {\tt update} calls.

By multiplexing control information for many sockets on each CM call,
the overhead from kernel crossings is mitigated at
the expense of managing more complicated data structures for the CM
interface.  Bulk querying is already performed in {\tt libcm} when
multiple flows are ready during a single {\tt ioctl} to determine
which flows can send data, but this completes the interface.

{\bf Trust issues.}  Because our goal was an architecture that did not
require modifications to receivers, we devised a system where
applications provide feedback using the {\tt cm\_update()} call.  The
consequence of this is that there is a potential for misuse, due to
bugs or malice.  For example, the CM client could repeatedly misinform
the CM about the absence of congestion along a path and obtain higher
bandwidth.  This does not increase the
vulnerability of the Internet to such problems, because such
abuse is already trivial.
More important are situations where users on the same
machine could potentially interfere with each other.  To prevent
this, the Congestion Manager would need to ensure that only
kernel-mediated (e.g. TCP) flows belonging to different users
can belong in the same macroflow.  Our current implementation does
not make an attempt to provide this protection.
Savage~\cite{Savage99b} presents several methods by which a
malicious receiver can defeat congestion control.
The solutions he proposes can be easily used with the CM;
we have already implemented byte-counting to prevent ACK division.

{\bf Macroflow construction}
When differentiated services,
or any system which provides different service to flows between
the same pair of hosts,
start being deployed, the CM would have
to reconsider the default choice of a macroflow.  We expect to be able
to gain some benefit by including the IP differentiated-services field
in deciding the composition of a macroflow.

Finally, we observe that remote LANs are not often the bottleneck for
an outside communicator.  As suggested in~\cite{rfc2140,Savage99}
among others, aggregating congestion information about remote sites
with a shared bottleneck and sharing this information with local peers
may benefit both users and the network itself.  A macroflow may thus
be extended to cover multiple destination hosts behind the same
shared bottleneck link.  Efficiently determining such
bottlenecks remains an open research problem.

{\bf Limitations}  
The current CM architecture is designed only to handle unicast
flows.  The problem of congestion control for multicast flows
is a much more difficult
problem which we deliberately avoid.  UDP applications using
the CM are required to perform their own loss detection,
requiring potential additional application complexity.
Implementing the Congestion Manager protocol discussed
in~\cite{Balakrishnan99a} would eliminate this need,
but remains to be studied.

%% file: related.tex
\section{Related work}
\label{s:related}

Designing adaptive network applications has been an active area of
research for the past several years.  In 1990, Clark and
Tennenhouse~\cite{Clark90} advocated the use of {\em application-level
framing} (ALF) for designing network protocols, where protocol data
units are chosen in concert with the application.  Using this
approach, an application can have a greater influence over deciding
how loss recovery occurs than in the traditional layered approach.
The ALF philosophy has been used with great benefit in the design of
several multicast transport protocols including the Real-time
Transport Protocol (RTP)~\cite{rfc1889}, frameworks for reliable
multicast~\cite{Floyd95,Raman98}, 
and Internet video~\cite{McCanne96,Rejaie99}.

Adaptation APIs in the context of mobile information access were
explored in the Odyssey system~\cite{Noble97}.  Implemented as a
user-level module in the NetBSD operating system, Odyssey provides API
calls by which applications can manage system resources, with upcalls
to applications informing them when changes occur in the resources
that are available.  In contrast, our CM system is
implemented in-kernel since it has to manage and share resources
across applications (e.g., TCP) that are already in-kernel.  This
necessitates a different approach to handling application callbacks.
In addition, the CM approach to measuring bandwidth and other network
conditions is tied to the congestion avoidance and control algorithms,
as compared to the instrumentation of the user-level RPC mechanism in
Odyssey.  We believe that our approach to providing adaptation
information for bandwidth, round-trip time, and loss rate complements
Odyssey's management of disk space, CPU, and battery power.

The CM system uses application callbacks or {\em upcalls} as an
abstraction, an old idea in operating systems.  Clark describes
upcalls in the Swift operating system, where the motivation is a lower
layer of a protocol stack synchronously invoking a higher-layer
function across a protection boundary~\cite{Clark85}.  The Mach system
used the notion of {\em ports}, a generic communication abstraction
for fast inter-process communication (IPC).  POSIX specifies a
standard way of passing ``soft real-time signals'' that can be used to
send a notification to a user-level process, but it restricts the
amount of data that can be communicated to a 32-bit quantity.

Event delivery abstractions for mobile computing have been explored
in~\cite{Badrinath95b}, where ``monitored'' events are tracked using
polling and ``triggered'' events (e.g., PC card insertion) are
notified using IPC.  This work defines a language-level mechanism
based on C++ objects for event registration, delivery, and handling.
This system is implemented in Mach using ports for IPC.

Our approach is to use a {\tt select()} call on a control socket to
communicate information between kernel and user-level.  The recent
work of Banga {\em et al.} \cite{banga+:usenix99}  to improve the performance of this type of
event delivery can be used to further improve our performance.

The Microsoft Winsock implementation is largely callback-based, but
here callbacks are implemented as conventional function calls since
Winsock is a user-level library within the same protection boundary as
the application~\cite{Quinn99}.  The main reason we did not implement
the CM as a user-level daemon was because TCP is already implemented
in-kernel in most UNIX operating systems, and it is important to share
network information across TCP flows.

Quality-of-service (QoS) interfaces have been explored in several
operating systems, including Nemesis~\cite{Leslie96}.  Like the
exokernel approach~\cite{Kaashoek97} and SPIN~\cite{Bershad+:sosp15},
Nemesis enables applications to perform as much of the processing as
possible on their own using application-specific policy, supported by
a set of operating system abstractions different from those in UNIX.  Whereas
Nemesis treats local network-interface bandwidth as the resource to be
managed, we take a more end-to-end approach of discovering the
end-to-end performance to different end-hosts, enabling sharing across
common network paths.  Furthermore, the API exported by Nemesis is
useful for applications that can make resource reservations, while the
CM API provides information about network conditions.  Some ``web
switches'' provide traffic shaping and QoS
based upon application information, but do not provide integrated
flow management or feedback to the applications creating the data.

Multiple concurrent streams can cause problems for TCP congestion
control.  First, the ensemble of flows probes more aggressively for
bandwidth than a single flow.  Second, upon experiencing congestion
along the path, only a subset of the connections usually reduce their
window.  Third, these flows do not share any information between each
other.  While we propose a general solution to these problems,
application-specific solutions have been proposed in the literature.
Of particular importance are approaches that multiplex several
logically distinct streams onto a single TCP connection at the
application level, including Persistent-connection HTTP
(P-HTTP~\cite{Padmanabhan94}, part of HTTP/1.1~\cite{rfc2068}), the
Session Control Protocol (SCP)~\cite{Spero96}, and the MUX
protocol~\cite{Gettys96}.  Unfortunately, these solutions suffer from
two important drawbacks.  First, because they are
application-specific, they require each class of applications (Web,
real-time streams, file transfers, etc.)  to re-implement much of the
same machinery.  Second, they cause an undesirable coupling between
logically different streams: if packets belonging to one 
stream are lost, another stream could stall even if none of its
packets are lost because of the in-order ``linear'' delivery forced by
TCP.  Independent data units belonging to different streams are no
longer independently processible and the parallelism of downloads is
often lost.

%% file: concl.tex
\section{Conclusion}
\label{s:concl}

The CM system enables applications to obtain an unprecedented degree
of control over what they can do in response to different network
conditions.  It incorporates robust congestion control algorithms,
freeing each application from having to re-implement them.  It exposes
a rich API that allows applications to adapt their transmissions at a
fine-grained level, and allows the kernel and applications to
integrate congestion information across flows.

Our evaluation of the CM implementation shows
that the callback interface is effective for a variety of
applications, and does not unduly burden the programmer
with restrictive interfaces.  From a performance standpoint,
the CM itself imposes very little overhead; that which remains
is mostly due to the unoptimized nature of our implementation.
The architecture of programs implemented using UDP imposes some
additional overhead, but the cost of using the CM after this architectural
conversion is quite small.

Many systems exist to deliver content over the Internet
using TCP or home-grown UDP protocols.  We believe that
by providing an accessible, robust framework for congestion control
and adaptation, the Congestion Manager can help
improve both the implementation and performance
of these systems.

The Congestion Manager implementation for Linux is available from
our web page, {\url{http://nms.lcs.mit.edu/projects/cm/}}.

\section*{Acknowledgements}
We would like to thank 
Suchitra Raman and Alex Snoeren
for their helpful feedback and suggestions; and
the Flux research group at the University of Utah
for providing their network testbed
\footnote{Supported by NSF grant ANI-00-82493, DARPA/AFRL grant F30602-99-1-0503 and Cisco}.  Finally,
we would like to thank our shepherd, Peter Druschel, and the
anonymous reviewers for their numerous helpful comments.
This work was supported by grants
from DARPA, IBM, Intel, and NTT Corporation.